\documentclass[12pt]{iopart}
\usepackage{graphicx}
\usepackage{iopams}
\usepackage{bm}
\begin{document}

\title{Interfacial depinning transitions in disordered
  media: revisiting an old puzzle}

\author{Bel\'en Moglia, Ezequiel V. Albano}
\address{Instituto de F\'{i}sica de L\'{i}quidos y
  Sistemas Biol\'{o}gicos (IFLYSIB),
  Universidad Nacional de La Plata, CONICET CCT-La Plata; Calle 59 Nro 789, (1900) La Plata, Argentina}
\address{Departamento de F\'{i}sica, Facultad de Ciencias Exactas,
  Universidad Nacional de La Plata, La Plata, Argentina}

\author{Pablo Villegas, Miguel A. Mu\~noz}
\address{ Departamento
  de Electromagnetismo y F{\'\i}sica de la Materia e Instituto Carlos
  I de F{\'\i}sica Te\'orica y Computacional. Universidad de
  Granada. Facultad de Ciencias.  E-18071, Granada, Spain}
  
  \date{\today}

\begin{abstract}
  Interfaces advancing through random media represent a number of
  different problems in physics, biology and other disciplines. Here,
  we study the pinning/depinning transition of the prototypical
  non-equilibrium interfacial model, i.e. the Kardar-Parisi-Zhang
  equation, advancing in a disordered medium. We analyze separately
  the cases of positive and negative non-linearity coefficients, which
  are believed to exhibit qualitatively different behavior: the
  positive case shows a continuous transition that can be related to
  directed-percolation-depinning while in the negative case there is a
  discontinuous transition and faceted interfaces appear. Some studies
  have argued from different perspectives that both cases share the
  same universal behavior.  Here, by using a number of computational
  and scaling techniques we shed light on this puzzling situation and
  conclude that the two cases are intrinsically different.
\end{abstract}
\maketitle

\section{Introduction}
The study and characterization of growing interfaces under
non-equilibrium conditions is a topic of interdisciplinary interest
\cite{HHZ,barabasi,spohn,krug}. Moving interfaces are often found in
physics (crystal and amorphous material growth, polymers and colloids,
granular matter, wetting, thin films), physical-chemistry (catalysis,
corrosion, reaction front propagation), biology (cellular, fungal, and
bacterial colonies growth, cell-sorting, wound healing, tumor
expansion), etc.  Understanding the properties of interfaces in
relation to phenomena such as corrosion, adhesion, wetting, friction,
micro- or nano-fluidics, etc. is essential for the development of
technological applications. Moreover, the study of interfaces is of
fundamental interest as a classical problem in statistical mechanics
as they constitute a canonical example of critical phenomena and
generic scale-free behavior in systems away from thermal equilibrium.

Within this broad context, the Kardar-Parisi-Zhang (KPZ) dynamics
\cite{kpz} represents the simplest and broadest universality class of
non-equilibrium growth \cite{HHZ,barabasi,spohn,krug}.  Its study has
been recently boosted by remarkable experimental and theoretical
breakthroughs
\cite{Take1,Take2,Johansson,Sasa,Calabrese,Prahofer,Canet,huergo2010,moglia2013}
which have triggered renewed interest. The KPZ interfacial dynamics is
defined by the Langevin equation
\begin{equation}
\label{kpz}
\partial_t h ({\bf x},t) = \nu {\bf \nabla}^2 h ({\bf x},t)+ 
\lambda (\nabla h ({\bf x},t))^2 + F + \eta({\bf x},t),
\end{equation}
where $h({\bf x}, t)$ is the local height of the interfaces, $F > 0$
is a driving force, $\eta({\bf x},t)$ is a zero-mean delta-correlated
Gaussian noise, the first term on the right-hand side (with
proportionality constant $\nu$) describes the relaxation of the
interface caused by the surface tension and, finally, $\lambda (\nabla
h)^2$ is the dominant nonlinear term. This last term accounts for lateral
growth and breaks the up-down symmetry in such a way that the
interface is not invariant under the transformation $h \rightarrow
-h$.  

Interfacial roughening properties are customarily analyzed by
measuring the {\em global} interface width:
\begin{equation}
\label{W}
W(L,t) = \langle[\overline{h(x,t) - \overline h]^2}\rangle^{1/2},
\end{equation}
\noindent where the overbar stands for spatial averages (in a system
of size $L$) and brackets denote disorder average. Usually, $W(L,t)$
obeys the Family-Vicsek dynamic scaling ansatz
\cite{family,HHZ,barabasi}, namely
\begin{equation}
\label{FV-globalwidth}
W(L,t) = t^{\alpha/z} f(L / \xi(t)),
\end{equation}
\noindent where the scaling function $f(u)$ obeys
\begin{equation}
\label{FV-forf}
f(u) \sim
\left\{ \begin{array}{lcl}
     u^{\alpha}     & {\rm if} & u \ll 1 \\
     {\rm constant} & {\rm if} & u \gg 1
\end{array}
\right.
\end{equation}
\noindent 
where $\alpha$ is the roughness exponent characterizing the stationary
(or saturated) regime, $\xi(t) \sim t^{1/z}$ is the correlation length
in the direction parallel to the interface, $z$ the dynamic
exponent, and $\beta= \alpha/z$ is the growth exponent that
governs the short-time behavior of the interface roughening. In
particular, for one-dimensional systems in the KPZ universality class $\alpha=1/2$,
$z=3/2$ and $\beta=1/3$, which have been measured in an overwhelming
variety of models and also experimentally
\cite{HHZ,barabasi,spohn,krug,Take1,huergo2010}.

Deviations from the previous values have also been reported in some
experimental set-ups, for which it can be argued that the interfacial
behavior is crucially affected by the presence of random pinning
forces, i.e. by quenched disorder or heterogeneity in the physical
background \cite{HHZ,barabasi}.  These situations can be addressed by
replacing the noise term $\eta({\bf x},t)$ in Eq.(\ref{kpz}) by a
quenched noised $\eta({\bf x},h)$, accounting for spatial (quenched)
heterogeneity
\begin{equation}
\label{qkpz}
\partial_t h(x,t)= \nu {\bf \nabla}^2 h(x,t) + 
\lambda (\nabla h(x,t))^2 + F + \eta({\bf x},h),
\end{equation}
\noindent
with $\langle \eta({\bf x},h) \eta({\bf x'},h')\rangle = \delta({\bf
  x}-{\bf x}') \Delta(h-h')$ (where $\Delta$ is some fast-decaying
function and $F$ is a external driving force), which is known as the
quenched Kardar-Parisi-Zhang (QKPZ) equation.  This equation is
usually complemented with the prescription that the interface is not
allowed to move backwards (i.e. $\partial_t h(x,t) <0
\rightarrow \partial_t h(x,t) =0$).  Equation (\ref{qkpz}) exhibits a
pinning/depinning phase transition at a certain critical value, $F_c$,
of the external driving force, $F$ \cite{HHZ,barabasi}: for $F > F_c$,
interfaces move with a finite velocity while for $F < F_c$ they
ineluctably become pinned by the impurities represented by the quenched noise.

Remarkably, the case in which the non-linearity acts in the same
direction as the driving force ($\lambda >0$) appears to differ
qualitatively from the one in which these two forces oppose each other
($\lambda <0$): for positive values of $\lambda$ (i.e. the positive
QKPZ or P-QKPZ equation) the depinning transition is smooth (second
order), while for negative $\lambda$ (i.e. the negative QKPZ or N-QKPZ
equation) it is abrupt (first order).  The underlying reason for such
a difference can be easily understood; taking Eq.(5) with quenched
noise, averaging over noise, integrating in $x$, and imposing the
stationary condition, one obtains
\begin{equation}
\lambda s^2 + F =0  
\end{equation}
where $s= \sqrt{\langle (\nabla h)^2 \rangle}$ is the average local
slope.  This equation has a non-trivial solution with $s>0$ if and
only if $\lambda <0$, corresponding to the pinned phase.  This
solution corresponds to faceted interfaces of average slope $s$ and
does not have a counterpart in the positive case, $\lambda
>0$.  Observe that the angle of the between facets, $\theta$, (see
Figure 1) obeys $s= \tan((\pi-\theta)/2) \propto 1/\sqrt\lambda$ and
reaches a maximum value at the depinning transition.

The faceted solution ceases to exist at $F=F_c$ where the interface
becomes depinned. Once the faceted solution breaks down, the interface
velocity $\langle \partial_t h \rangle$, experiences a first-order
transition and jumps from $0$ to some constant stationary value. Even
if the transition is discontinuous, the interface shows aspects of
scale invariance both above and below the transition point.  This type
of hybrid situations sharing aspects of first order transition and
scale invariance is known in the literature (see e.g. \cite{Liggett}).

Even if this simple argument suggests that the positive and negative
cases should exhibit intrinsically different features, a
renormalization group calculation reveals no difference between the
positive and the negative cases \cite{RG}. Indeed, the renormalized
value of $\lambda^2$ diverges, suggesting the existence of a strong
coupling fixed point for any value $\lambda \neq0$.  The renormalized
value of $\lambda^2 $ was measured in simulations of the N-QKPZ,
revealing that it does not diverge but stays finite even as the system
approaches its critical point, suggesting that the renormalization
group calculation might break down in this case. But he situation at
this theoretical level has not been clarified thus far.

From the computational side, the QKPZ dynamics has been profusely
studied both for positive and negative non-linearities in one spatial
dimension. Tang et al. \cite{tango} proposed that the P-QKPZ equation
can be effectively described by the statistics of disorder pinning
paths and, hence, mapped into the so-called directed percolation
depinning (DPD) model \cite{DPD}. Thus, the roughness exponent is
given by the ratio of the two correlation length exponents, in the
parallel and perpendicular direction of the directed percolation
cluster, namely $\alpha = \nu_{\bot}/\nu_{||}$ ($\simeq 0.63$);
similarly it follows that $z=1$ and hence $\beta=\alpha$.  These results
agree with numerical simulations of systems in this class
\cite{les,Kim}.  On the other hand, numerical studies of different
models with efective negative non-linearity confirmed the formation of
facets and the existence of a jump at the transition
\cite{jeong1,jeong2,cantabros}.

Self-organized models --in which interfaces self-tune to the
transition point \cite{Paths}-- have also been proposed and studied in
this context.  Sneppen \cite{Sneppen} proposed two different
self-organized growth models in random media one leading to facets and
the other not and concluded that one lies in the N-QKPZ class while
the other behaves as P-QKPZ.  On the contrary, Choi et al. \cite{choi}
formulated two other similar self-organized models --with positive and
negative non-linearities respectively-- and concluded that the sign of
the non-linear term does not affect the universality class.

Aimed at clarifying this very confusing state-of-affairs, here we
revisit the P-QKPZ and the N-QKPZ equations.  Among other methods, we
analyze the results by employing spectral techniques to establish
whether the formation of facets --and ultimately the sign of the
non-linearity in the QKPZ equation-- plays a relevant role or whether
it does not.

\section{Anomalous scaling}
In some interfacial problems it is important to distinguish between
global and local roughening properties.  The {\em local} interface
width $w(l,t)$ is defined as
\begin{equation}
\label{W-local}
w(l,t) = \langle[\overline{h(x,t) - \overline h]^2}\rangle^{1/2} , 
\end{equation}
\noindent where $\langle \cdots \rangle$ denote disorder average 
and the overbar an average over $x$ in windows of size $l$, obeying
\begin{equation}
\label{FV-localwidth}
w(l,t)=t^\beta f_A(l/\xi(t)),
\end{equation}
\noindent where $\beta$ is the growth exponent. Now the scaling function maybe  anomalous, i.e.
\begin{equation}
\label{f_A}
f_A(u) \sim
\left\{ \begin{array}{lcl}
     u^{\alpha_{loc}}     & {\rm if} & u \ll 1 \\
     {\rm const} & {\rm if} & u \gg 1, 
\end{array}
\right.
\end{equation}
\noindent where $\alpha_{loc}$ is a new independent exponent called
the local roughness exponent which is in general does not need to
coincide with its global counterpart, $\alpha$.

Ramasco et al. introduced a general dynamic scaling
ansatz for roughening interfaces which includes all the
previously-known forms of dynamic scaling as particular cases
\cite{cantabros} (see also \cite{c2,c3,c4}).
Implicit to this general scaling ansatz is the hypothesis that the
interface may exhibit two different types of behavior at short and
long scales respectively. The analysis relies on the structure factor
or power spectrum $S(k,t)$)
 \begin{equation}  
\label{S}
S(k,t) = 
\left\langle  \left| \frac{1}{\sqrt{L}} \int_{0}^{L}  dx h(x,t) e^{-ikx} 
  \right|^{2}   \right \rangle,
\end{equation}
\noindent where $k = 2\pi n/L$, with $n = 1, 2,....,L -1$. The generic
scaling ansatz for $S(k,t)$ proposed in \cite{cantabros} is 
\begin{equation}
\label{Skt}
S(k,t) =k^{-(2\alpha+1)} s(kt^{1/z}),
\end{equation}
\noindent with
\begin{equation}
\label{Anom-s}
s(u) \sim  
\left\{ \begin{array}{lcl}
 u^{2\alpha + 1} & {\rm if} &  u \ll 1 \\
     u^{2(\alpha-\alpha_s}) & {\rm if} &  u \gg 1,
     \end{array}
\right. 
\end{equation}
\noindent where $\alpha_s$ is the {\em spectral} roughness
exponent. If $\alpha_s\neq \alpha$ there is anomalous scaling, while
if $\alpha_s=\alpha$ the standard Family-Vicsek scaling is
recovered.  Remarkably, a novel type of anomalous scaling behavior
(with $\alpha = \alpha_{loc}=1$ and $\alpha_s > \alpha_{loc}$) was
theoretically predicted in \cite{cantabros}, and one of the previously
mentioned models by Sneppen (the one with facets) was argued to lie in
this family.

Let us remark that --as emphasized by Ramasco and coworkers
\cite{cantabros}-- $\alpha_s$ does not explicitly appear in the
scaling behavior of either $W(L,t)$, $w(l,t)$ or the height-height
correlation function $G(l,t)$ and, thus, can not be deduced from
measurements of these quantities, suggesting that a sound study of the
roughening properties should include spectral analyses.

\section{Results}
We solved numerically Eq.(\ref{qkpz}) with both positive and negative
non-linearities in one dimensional lattices and study its spectral
properties.  For that, we consider a standard finite-differences
discretization scheme for Eq.(\ref{qkpz}) in rings of size $L$ (i.e.
periodic boundary conditions are assumed)
\cite{barabasi,les,jeong2}. More refined algorithms as the one
proposed in \cite{Shim} could be implemented, but they are not
necessary for our purposes here.  Time is discretized in units of
$\Delta t = 0.01$, $\nu=1$, and --following previous analyses
\cite{jeong2}-- noise is taken to be uniformly distributed in $[-a/2,
a/2]$ with $a = 4.642$. Initial conditions correspond to a flat
interface $h(x,t=0) = constant$.  A fresh value of the quenched random
force is extracted at position $x$ whenever the interface advances at
such point; this value is kept fixed until the interface moves forward
again.  Ensemble averages are performed over at least $1000$ different
realizations of the quenched randomness.  Results have been verified
to be robust against changes in these choices.

\subsection{$\lambda >0$ (P-QKPZ)}
Figure \ref{profiles}-(a) shows interface profiles for the P-QKPZ case
(with $\lambda=0.5 > 0$ and $F=1$): the interface grows until it
becomes eventually pinned for $F<F_c$. The measured roughness exponent at the
transition point is $\alpha=0.63(1)$ in good agreement with the
expectation for the DPD class. Given that the universality of this
class is well understood \cite{HHZ,barabasi}, we have not performed
further extensive numerical studies of this positive $\lambda$ case.

\subsection{$\lambda <0$ (N-QKPZ)}
\begin{figure}
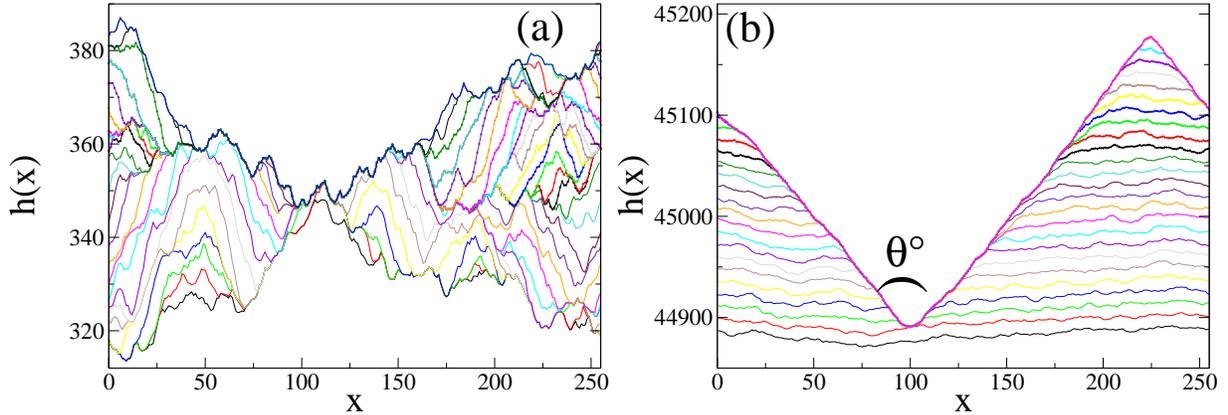

\includegraphics[width=8cm]{Figure1A.eps} \includegraphics[width=8cm]{Figure1B.eps}
\caption{Time evolution of a KPZ interface moving in a $(1 +
  1)-$dimensional disordered medium (system size $L = 256$). (a) {\bf
    P-QKPZ} case with $\lambda = 0.5$, $F = 1.00 < F_{c}$) and
  different times (from botton to top $t=199000$ to $t=232000$ in
  uniform intervals). (b) {\bf N-QKPZ} case with $\lambda = - 0.5$ and
  $F = 1.90 < F_{c}$) for different times (from botton to top
  $t=4030000$ to $t=4055000$ in uniform intervals).  The average angle
  at the bottom of the valley, $\theta =49(2)^{\circ}$, was obtained
  by averaging over $100$ different pinned interfaces.}
\label{profiles}
\end{figure}

Figure \ref{profiles}-(b) shows a profile in the N-QKPZ case ($\lambda
= -0.5 <0$) obtained close to the transition point $F_c \approx 1.98$.  Observe the
distinct shape of pinned interfaces exhibiting --as expected--
characteristic facets. In agreement with previous findings, we observe
a first-order pinning-depinning transition at which the averaged
interfacial velocity jumps discontinuously from zero to some positive
constant value.

\subsubsection{The depinned phase}
\begin{figure}
\centerline{
\includegraphics[width=8cm]{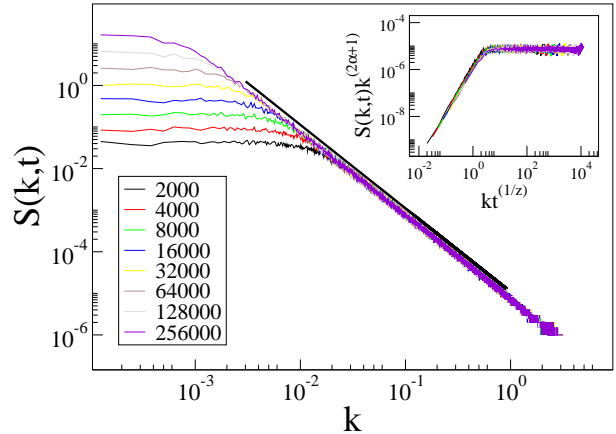}
}
\caption{Supercritical behavior in the N-QKPZ case: Double logarithmic
  plots of the structure factor, $S(k)$ versus the wavenumber $k$
  obtained for different times for $F = 3 \gg F_{c}$, $ L = 50000$ and
  averaging over $250$ configurations. The continuous straight line is
  a fit of the long-$k$ regime and has slightly been shifted up for
  the sake of clarity. It has a slope $-2.03(4)$ yielding $\alpha =
  0.51(2)$.  The inset shows a data collapse obtained using equation
  (\ref{Anom-s}) with $\alpha = 1/2$ and $ z = 3/2$.
\label{fig2}}
\end{figure}

For sufficiently large driving forces --deep into the depinned or
moving phase-- quenched disorder should be irrelevant above some
length and time scales, and the freely moving interface should
therefore follow standard KPZ dynamics. Indeed, taking $F = 3 \gg
F_{c}$ (cf. Figure \ref{fig2}) we find that $S(k,t)$ scales in the
large time regime scales as a power law with exponent $2\alpha + 1 =
2.03(4)$, i.e. with $\alpha = 0.515(20)$, as corresponds to standard
non-anomalous Family-Vicsek behavior (see the collapse obtained in the
inset of Figure \ref{fig2} with $\alpha = 1/2$ and $z =
3/2$). Therefore, the moving interface belongs to the standard KPZ
universality class, as expected.

\subsubsection{The pinned phase}
\begin{figure}[h]
\centerline{\includegraphics[angle=-90,width=8cm]{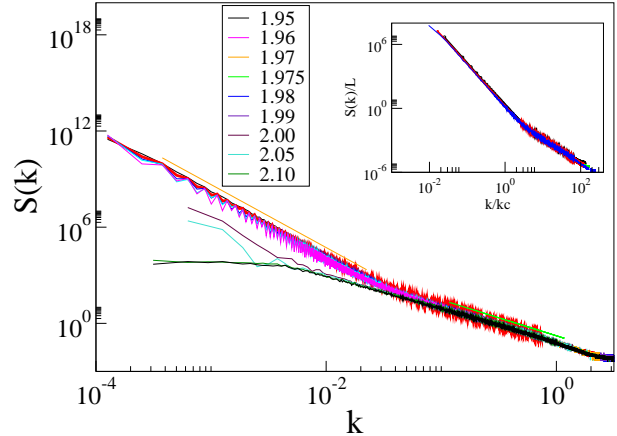}}
\caption{Double logarithmic plot of the structure function, $S(k)$ for
  the N-QKPZ case for both subcritical and supercritical values of $F$
  (system size $ L = 5000$).  The continuous straight lines are the
  fits of the large-$k$ regime (slope $-2.1(2)$, i.e. $\alpha =
  0.55(5)$) and the small-$k$ regime for subcritical forces (slope $ -
  3.99(2)$, i.e. $\alpha_s = 1.49(2)$), respectively. The fits have
  slightly been shifted up for visual clarity.  Some supercritical
  values of $F$ have been included in the plot to illustrate that the
  short-scale behavior is indistinguishable in both cases, and
  compatible with $\alpha=1/2$.  Inset: Log-log plots of the structure
  factor in the subcritical regime, rescaled with system size, versus
  $k/k_c(L)$ --where $k_c(L)$ is the value of $k$ at which the
  crossover occurs-- obtained for samples of different side $L$,
  i.e. $ L=10000, L=20000, L=50000, L=100000, L=200000$, and for $F =
  1.95 < F_{c}$. A nice curve collapse is observed.  }
\label{fig3}
\end{figure}

More interesting is the behavior of $S(k)$ for stationary pinned
interfaces, $F < F_{c}$. Figure \ref{profiles}-(b) shows results for a
single realization; it illustrates the development of a (single) well
defined pinning center close to $x = 100$ at which the interface
becomes eventually fully pinned. A careful inspection of Figure
\ref{profiles}-(b) reveals that the slopes around the peak are not
just straight lines but they have some intrinsic roughness. Therefore,
two different regimes are expected to emerge when computing the
structure function, corresponding to linear slopes and fluctuations on
top of them, respectively. This suggests the existence of anomalous
scaling. Indeed, as shown in Figure \ref{fig3}, $S(k)$ exhibits a
crossover between short and large $k$ regimes at a certain crossover
value, $k_{c}$.  Observe that, as illustrated in the inset of Figure
\ref{fig3}, the crossover between short and long scales is rather
insensitive to changes in $F$ and in $L$, revealing the absence of a
diverging correlation length.

The structure function of pinned interfaces (cf. Fig. \ref{fig3})
clearly shows two well separated regimes; the small-$k$ (large
wavelength) limit describes facets while the large-$k$ (short
wavelengths) corresponds to the fluctuations existing on the top of
the two facets.  From the slopes of the curve shown in Figure
\ref{fig3}, we obtain $\alpha_{s} = 1.49(2)$ in the small-$k$ regime,
i.e.  for the macroscopic faceted structures. Let us remark, that for
the trivial case of a perfectly faceted interface formed by identical
segments it is not difficult to show that the spectral roughness
exponent is $\alpha_{s} = 3/2$ \cite{cantabros}.  On the other hand,
we measure, $\alpha = 0.55(5)$ for the large-$k$ (small wavelength)
regime, which corresponds to the roughness that ``modulates'' the
slopes of the facets. This value is compatible with $\alpha=1/2$, as
obtained for depinned interfaces.

\begin{figure}
\centerline{
\includegraphics[width=8cm]{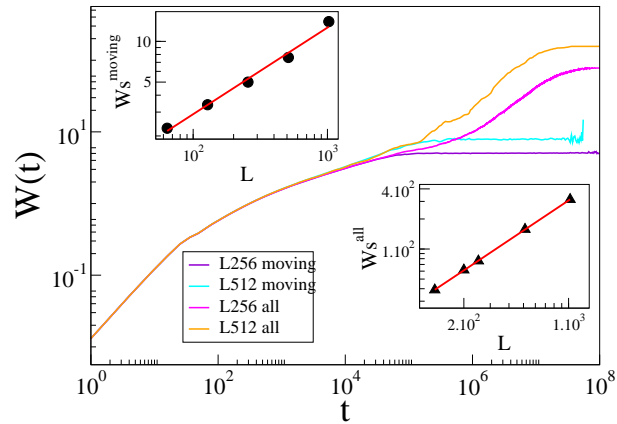}
}
\caption{Global interfacial width in the N-QKPZ case.  (a) Log-log
  plots of the global width $W$ versus $t$ obtained for $F = 1.90 <
  F_{c}$ and samples of different size, averaged over {\it all}
  realizations or restricted to {\it moving} interfaces (two curves,
  corresponding to $2$ different sizes are represented).  Upper inset:
  Log-log plot of the (stationary) saturation value of the global
  width $W$ versus sample size, $L$, for moving interfaces obtained
  for $5$ different system sizes (including the 2 sizes in the main
  plot).  The best fit of the straight line yields $\alpha^{moving}
  \approx 0.53$.  Lower inset: as the upper inset, but averaging over
  all runs (pinned and moving; the best fit gives $\alpha^{all} =
  1.003(8)$. }
\label{figtodos}
\end{figure}

\subsection{Global and local roughening}
Now we present results obtained by standard measurements of the global
and local interface roughness, (Eqs. (\ref{W}) and (\ref{W-local}),
respectively).  Figure \ref{figtodos} shows log-log plots of the
global interface width versus time, obtained for $F = 1.90 < F_{c}$.
Two types of averages are presented, either over all runs (labelled
{\it all}), or restricting the average to moving interfaces (label
{\it moving}). Observe that averages including all runs (and thus,
pinned faceted interfaces) have a larger roughness.

The roughness exponents corresponding to the global width measured for
depinned interfaces, $\alpha^{moving} \approx 0.53$, is consistent
with the value obtained for the large-$k$ regime of the structure
factor. Thus, the global width of moving interfaces captures the
roughness that ``modulates'' the slopes of the facets. On the other
hand, once pinned (i.e. faceted) interfaces are taken into account, we
obtain $\alpha^{all} \approx 1$, implying that the scaling is
dominated by linear facets.

\begin{figure}
\vspace{1.0cm}
\centerline{
\includegraphics[width=8cm]{Figure5A.eps}
\includegraphics[width=7.3cm]{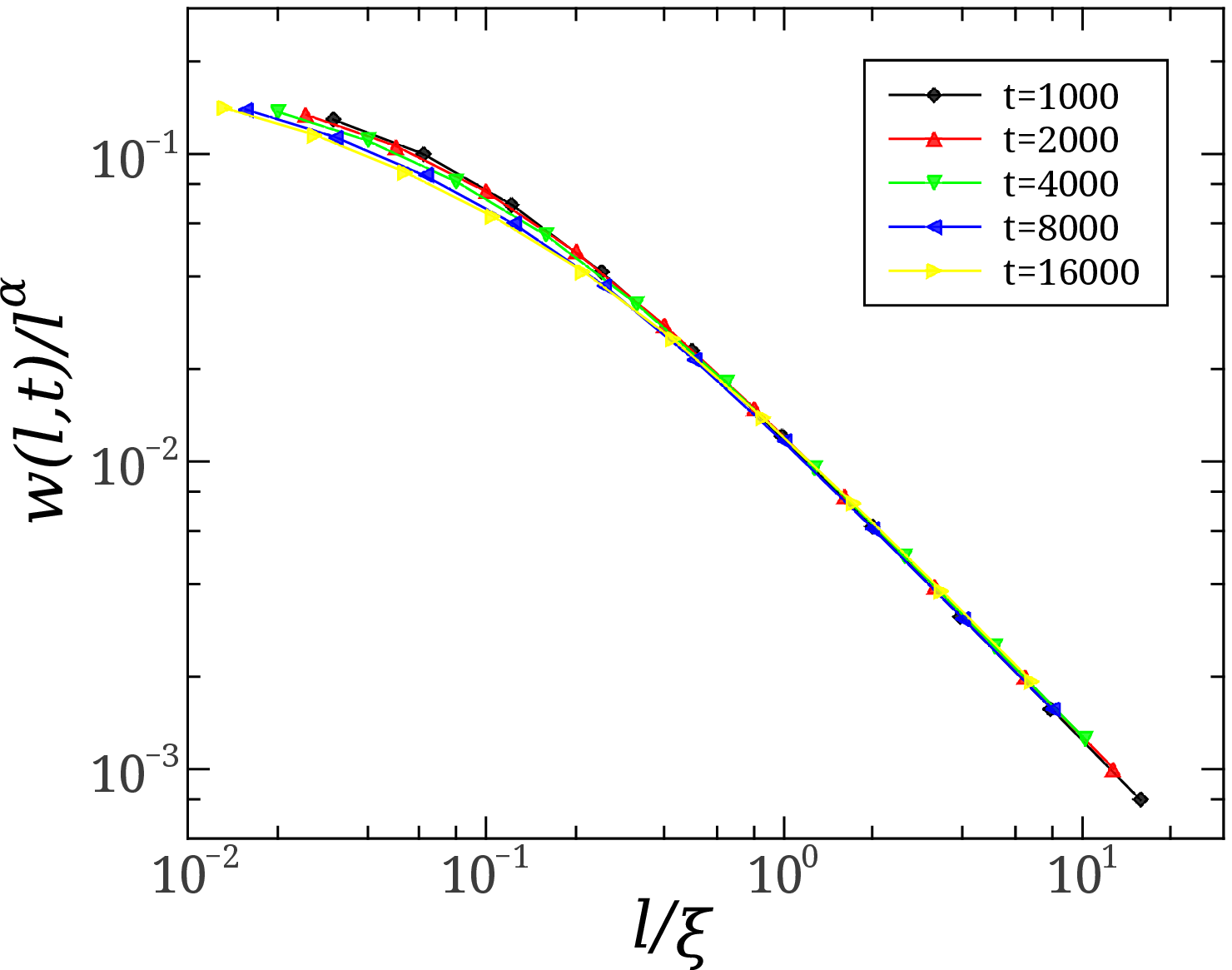}
}
\caption{Left: Local interfacial width in the N-QKPZ case. Log-log
  plots of the local width of the interface $w(l,t)$ versus $l$
  obtained for samples of size $L = 4096$ and different measurement
  times, as indicated, in the pinned phase ($F = 1.90 < F_{c}$ and
  averages over $500$ configurations).  Initially interfaces are flat
  and then, progressively, roughness develops.  Diamonds stand for
  pinned interfaces and the dashed line (which has been shifted for
  the sake of clarity) shows the best fit, corresponding to
  $\alpha_{loc} = 0.997(5)$.  Observe that the range in which the
  linear scaling can be observed grows as time increases and facets
  develop. Right: curve colapse obtained using the scaling form
  $w(l,t) \sim l^\alpha F(l/\xi)$ for times up to $t=16000$; for
  larger times a saturation length $\xi$ cannot be properly measured.}
\label{fig5}
\end{figure}

Figure \ref{fig5} shows log-log plots of the local width $w(l,t)$
(cf. Eq.  (\ref{FV-localwidth})) versus $l$ obtained for different
times. Measurements performed for pinned interfaces (in the $t
\rightarrow \infty$ limit) allow us to determine $\alpha_{local} =
0.997(5)$, confirming that for pinned interfaces both the local and
the global roughness exponents are asymptotically controlled by the
faceted structure.  On the other hand, employing the scaling form
$w(l,t) \sim l^\alpha F(l/\xi)$ where $F$ is a scaling function and
$\xi$ is a saturation or correlation length (i.e. the value of $l$
above which a constant local width is measured), and using $\alpha=1$
we obtain a good collapse as illustrated in the right of Figure 5 (see also
similar scaling laws for the pinned and depinned phases, in \cite{EPL}).

\subsection{Direct analysis of local fluctuations modulating facets}
Figure 6a shows a snapshot of a pinned configuration; the slopes of
the faceted structure have been fitted by two straight lines.  On top
of these linear structure there are fluctuations, as illustrated in
the inset of Fig6a, where the averaged slope has been locally
subtracted.  By computing the variance (R) around the linear fits for
facets of different linear size, we obtain the local width as a
function of the facet linear size, $l$ (see Fig.6b). It follows that the data
can be very well-fitted in a double-logarithmic plot by a
straight line with slope $0.51(1)$, suggesting again a local roughness,
compatible with $\alpha=1/2$.
\begin{figure}[h!]
\begin{centering}
\includegraphics[scale=0.52]{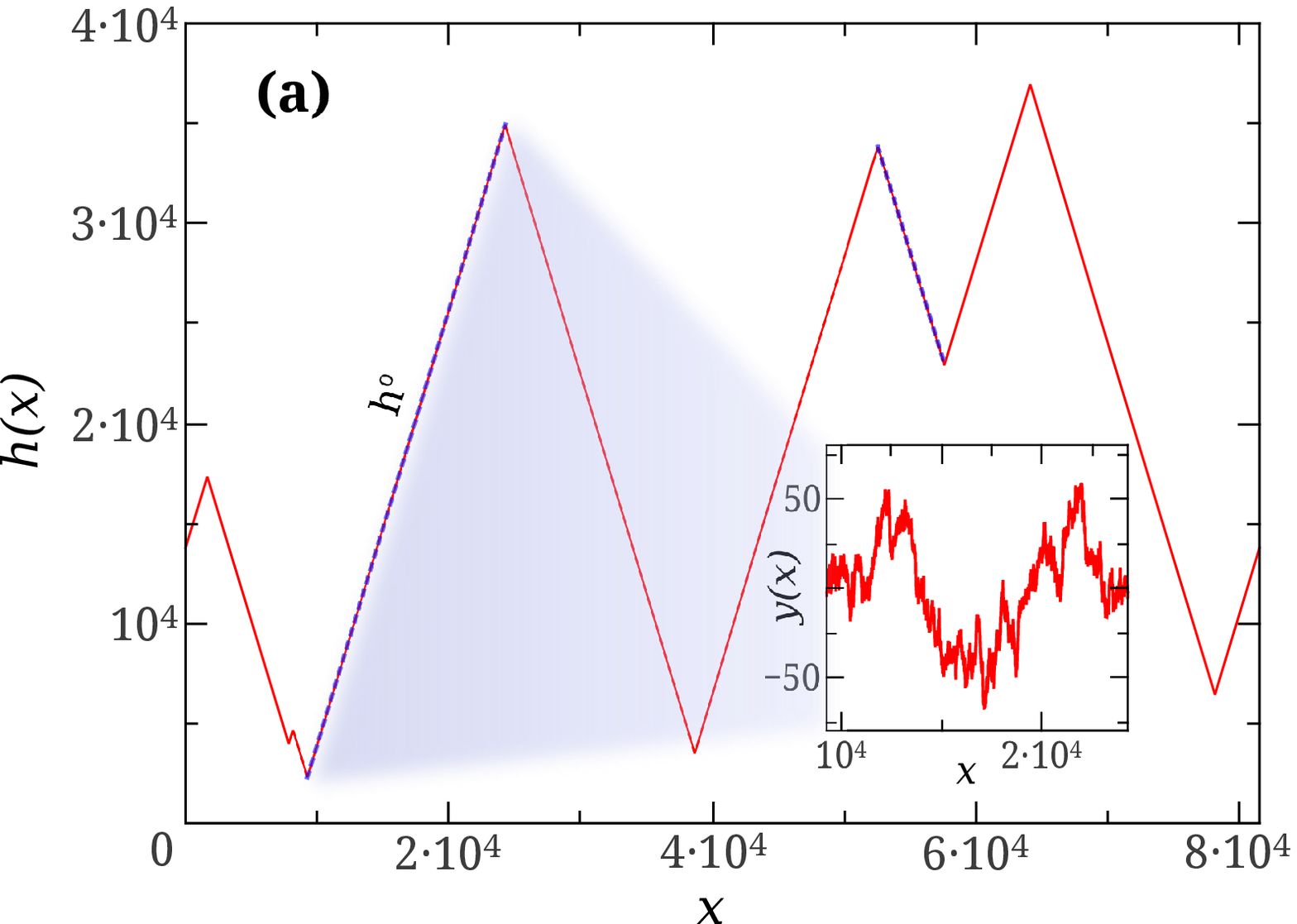}
\end{centering}
\begin{centering}
\includegraphics[scale=0.51]{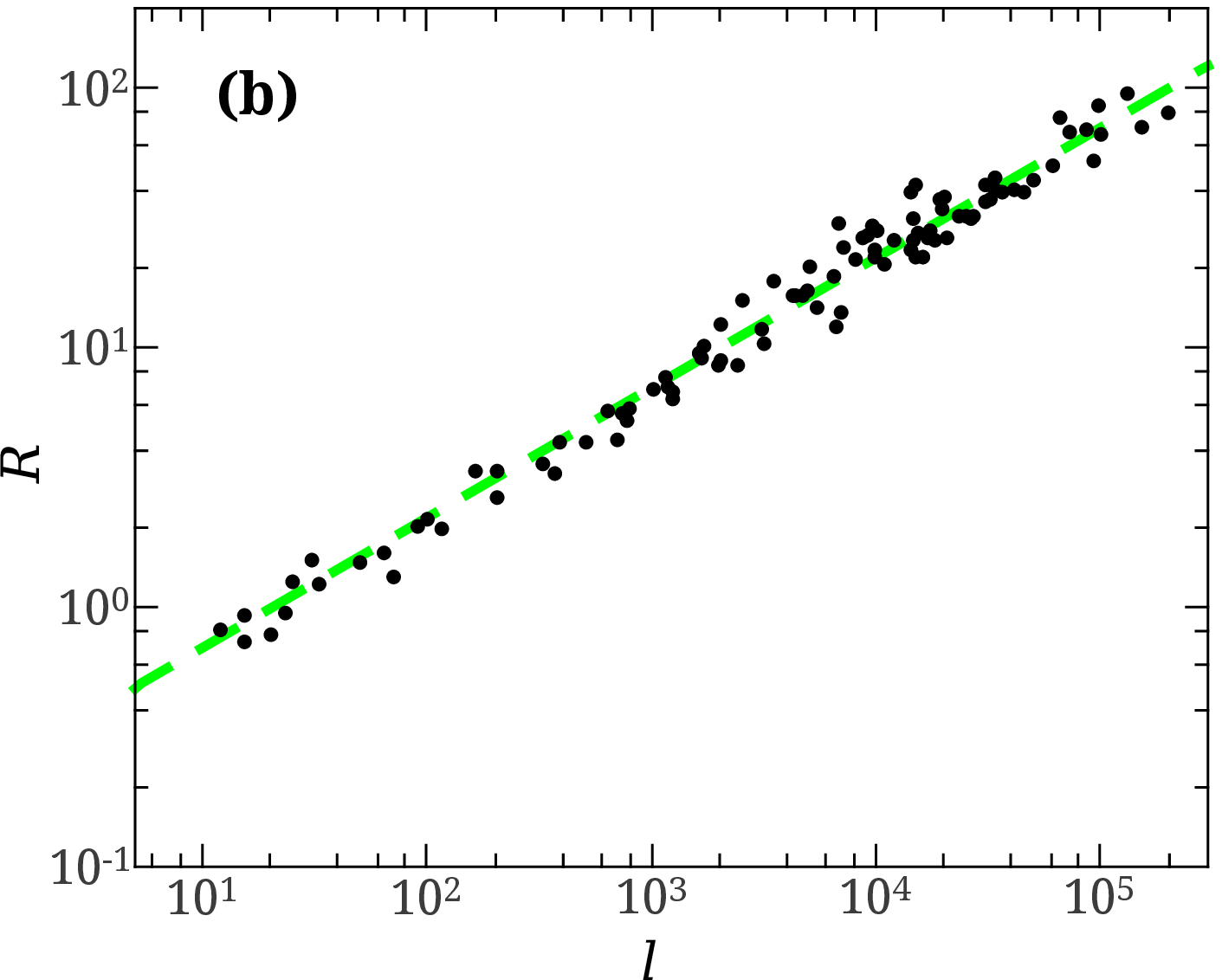}
\end{centering}
\caption{Analysis of local scale fluctuations in the N-QKPZ.  (a)
  Example of a pinned interface with $\lambda=-0.5$ and $F=1.94$.  The
  slopes of the facets can be linearly fitted (straight dashed lines
  of slope $h^{o}$), allowing us to estimate the slope and the mean
  squared error $R$ around it. Inset: zoom of the local fluctuations
  $y(x)$ around one of the facets.  (b) Log-log plot of
  $R\,\left(w_{local}\right)$ versus the linear size of the facets
  (lines are guides to the eye).  The best fit is obtained for a local
  roughness exponent $0.51(1)$, close to the KPZ value $\alpha=1/2$.}
\end{figure}

\section{Discussion and Conclusions}
We have presented a full characterization of the interfacial growing
behavior of the KPZ equation with quenched noise and a negative value
of the coefficient in the nonlinear term (cf. Eq. (\ref{qkpz})).  The
positive case exhibits a continuous phase transition in the DPD universality class,
while in the negative case we have found evidence of a discontinuous transition
separating a pinned phase, characterized by faceted interfaces and a moving
KPZ-like phase. Our study is focused on the negative case and our main conclusions are: 
\begin{enumerate}

\item Measurements of the structure factor of pinned interfaces show
  anomalous scaling behavior that can be considered as a particular
  case of the general scaling theory proposed by Ramasco et al. as
  applied to pinned interfaces (i.e. with no explicit time
  dependence). $S(k)$ exhibits a crossover between the small-$k$
  regime with $\alpha_{s} \approx 1.5$ (controlled by facets) and the
  large-$k$ regime with $\alpha \approx 0.55$.

\item
Standard measurements of the local and global widths and the analysis
of its scaling behavior within the pinned phase ($F < F_{c}$) yield
$\alpha_{local} \approx \alpha^{all} \approx 1$. However, by excluding
pinned (faceted) interfaces in the calculation of the average we obtained
$\alpha^{moving} \approx 0.53$, consistent with the large-$k$ scaling
of the structure factor.

\item Finally, direct measurements of the fluctuation around the
  facets reveal that local fluctuations can be well represented by a
  roughening exponent $\alpha \approx 0.51$.

\end{enumerate}

All these results taken together suggest that local roughening is
controlled by the standard KPZ roughening exponent. This result is in
agreement with the finding in \cite{Szendro} for a similar interfacial
model with columnar disorder (i.e. $\eta=\eta(x)$); this model was
reported to exhibit facets which roughness profiles on top of them,
controlled by a $0.5$ exponent. Furthermore, in this same work
\cite{Szendro}, the authors showed analytically that the dynamics of
facets can be decoupled from short scale fluctuations, and that these
latter ones exhibit KPZ roughness.  An almost identical calculation
leads us to the same conclusion here: local and global dynamics are
decoupled; on the one hand there are facets and on the other there are
short-scale KPZ-like fluctuations. 

Therefore, we have not found any evidence of a continuous transitions
nor of roughness exponents around $0.63$, characteristic of the DPD
class in the negative case, and we can safely conclude that the two
cases, with positive and negative non-linearities are clearly
different. Obviously, the origin in this difference stems from the
facet formation in the negative case; thus it would be reasonable to
conjecture that by running simulations in tilted systems --with a tilt
equal or larger to the critical slope-- there should not be an abrupt
transition between faceted and non-faceted/moving interfaces. One should
not observe a continuous transition and exponent values at
the transition point compatible with DPD class, as indeed numerically
verified in \cite{jeong1}.

Beside of this new study, some important questions remain unsolved and
the study of interfaces in random media remains an intriguing research
area. For example, analyzing in detail what happens in
physically-more-relevant higher dimensional systems (e.g. in two
dimensions) where pinning paths (and thus DPD) are expected to be
replaced by ``pinning surfaces'' \cite{surfaces} is left for a future
work.

Interestingly, a similar physical situation arises in the study of KPZ
interfaces bounded by a wall which is relevant in the study of
non-equilibrium wetting \cite{wetting, wetting2} and synchronization transitions
\cite{synchro}. Under these circumstances, the case $\lambda >0$ has
been shown to be radically different from the $\lambda <0$ one; the
corresponding associated problems have very different physical
behavior and they belong to two distinct universality classes
\cite{MN}. Therefore, it seems that under diverse circumstances,
positive and negative KPZ non-linearities describe very different
situations.

\vspace{0.5cm}

{\bf Acknowledgments:}
We acknowledge financial support from Acci\'on Integrada
hispano-argentina, AR2009-0003; MAM acknowledges support from J. de
Andaluc\'{i}a project of Excellence P09-FQM-4682 and from the Spanish
MEC project FIS2009--08451.  B.M and E.V.A acknowledge the financial
support of CONICET (PIP--0143) and UNLP (Argentina). We are thankful to F. de los
Santos, J.A. Bonachela, and J.M. L\'opez for useful discussions and/or
a critical reading of the manuscript.

\vspace{0.65cm}

{\bf {References}}


\end{document}